\documentstyle{article}
\begin{document}
\title{ Neural network learning dynamics in a path integral framework } 
\author{ 
J.Balakrishnan \thanks{E-mail : janaki@serc.iisc.ernet.in, ~janaki@hve.iisc.ernet.in } \\ 
Department of High Voltage Engineering, Indian Institute of Science,\\ 
Bangalore -- 560 012,
India. }                                                                
\date{\mbox{\ }} 
\maketitle 
\vspace{1.5cm}
\begin{center} 
P.A.C.S. number ~~: ~~ 02.90.+p, ~05.90.+m , 87.10.+e ~~~~~~
\end{center} 
\vspace{1.5cm} 
\begin{abstract}

A path-integral formalism is proposed for studying the dynamical evolution 
in time of patterns in an artificial neural network in the presence of noise. 
An effective cost function is constructed which determines the unique global 
minimum of the neural network system. The perturbative method discussed also 
provides a way for determining the storage capacity of the network. 
\end{abstract}
\newpage 

\subsection*{1. ~Introduction}

It has been of interest since long to understand the mechanism of learning 
and memory in biological systems and machines. Studies of associative memory 
have sought to model the process of pattern recognition and recall using 
specific cost functions.\\
The similarity of the McCulloch-Pitts neural network to the Ising spin system 
has enabled statistical physics approaches [1-3,4] to be used to get information 
like the storage capacity of the network. 
In the ``space of interactions'' approach of Gardner et al [2,3], an imbedding 
condition is postulated and the energy function counts the number of 
weakly-imbedded pattern spins which have stability less than a specified 
value. Though this approach does a systematic study of a neural network 
configuration at any given instance of time, it does not address the 
question of time evolution of the network configuration. 
Some years ago, Hertz et.al.[5,6] studied the dynamics of learning in a  
single-layer neural network 
using a Langevin equation for the evolution 
in time of the synaptic efficacies. 
In these papers the authors have investigated the role of noise in 
learning and studied the possible phase transitions in the learning process.\\

In our work we have taken such a viewpoint, of looking at the learning 
process as a non-equilibrium stochastic process, as our starting 
point for constructing a path-integral framework for studying neural network 
dynamics.\\ 

The problem of neural networks getting trapped into spurious states or local 
minima is well known and a method for directly getting to the global minimum 
of the network is highly desirable. Of much more interest is a systematic 
theory which gives a framework for determining the global minimum of the 
neural network model, independent of the choice of the cost function.\\  
An attempt has been made in this work to achieve this 
through the path-integral framework using concepts from quantum and 
statistical field theory --- we have considered a perceptron only for 
the sake of simplicity.  

In this framework it is seen that the patterns in the network 
settle from non-equilibrium states into certain attractor states which 
correspond to those with lowest energy at equilibrium, or for large values 
of the time. We construct an ``effective cost function'' for any cost 
function one starts with, and discuss how the global minimum for a 
neural network can be determined and how one can calculate the storage 
capacity of the network with this construction.\\

\subsection*{2. ~Langevin dynamics in a path-integral approach } 
Langevin dynamics has been applied to analyse disordered systems, in particular, to spin glasses and to retrieval processes in attractor neural network models with fixed weights, by a number of authors [7]. In our work, we look at learning dynamics from a slightly different viewpoint --- through a path-integral framework, and with a dynamical evolution in time for the synaptic efficacies. \\ 
As in [5,6], we view the problem of learning in neural networks 
as a stochastic process and for simplicity, we look at the 
perceptron only, with one layer of connections. We 
postulate a stochastic Langevin dynamics for 
the evolution in time of the synaptic efficacies $w_{ij}$. 
\begin{equation} 
\frac{\partial \omega_i(\tau)}{\partial \tau} = - \gamma T\frac{\delta E( 
\omega_i)}{\delta \omega_i(\tau)} + \eta_i(\tau) 
\end{equation} 
where the input index $i$ takes values from 1 to N and we have omitted the 
output index which can be treated separately. $\eta_i(t)$ stands for a random  
white noise source, $T$ is the noise level, $E$ is the cost function and 
the parameter $\gamma$ 
describes the learning rate. The system can alternatively be thought to 
be coupled to a heat reservoir at 
temperature $T$ (which represents the noise level) and evolves in time 
$t$ until it reaches an equilibrium 
configuration at $t \rightarrow \infty$ . \\ 
Here, and in the following, the space label has been suppressed and 
the index $i$ should be interpreted to include the space variable also.
We assume that the noise sources are gaussian with the correlations :\\  
\begin{eqnarray} 
\langle \eta_i(\tau) \rangle &=& 0 \nonumber\\
\langle \eta_i(\tau)\eta_j(\tau') \rangle &=& 2\gamma T \delta_{ij}
\delta(\tau - \tau') 
\end{eqnarray} 
The values of the synaptic efficacies at any instant of time are determined 
by solving (1) and the correlations between them can be calculated using 
(2).\\  
We wish to construct a partition function and a path integral framework 
for this simplest 
type of neural network which incorporates the time evolution of the 
synaptic strengths. 
Pattern recall and recognition takes place best when the Hamming distance 
between the target $\zeta^\mu_i$ for pattern $\mu = 1,\dots, p $ 
and the input pattern $\xi^\nu_i$ is minimal for $\mu = \nu$. 
This means that the synaptic strengths $\omega_{ij}$ change in such a way 
as to descend the cost function surface. \\

In an artificial neural network, the problem of the system getting trapped in
its various local minima or spurious states is a familiar one. 
Here, prior knowledge of the global minimum for the system and the 
synaptic efficacies corresponding to it would be very useful as it would save a 
great deal of effort and computer time and enable greater efficiency in 
solving pattern recognition and associative memory problems. \\ 

We adapt the procedure which was elucidated by Gozzi [8] 
in a different context, 
to the neural network system and write down a partition function for 
this system evolving in time through the Langevin dynamics of (1) :\\  
\begin{equation} 
Z[J] = {\cal N} \prod_{i,\mu}^{ } \int_{0}^{\tau} D\omega_i D\eta_i 
e^{-\frac{1}{\gamma T} \int 
J_i(\tau')\omega_i(\tau')d\tau'} P(\omega(0)) \delta(\omega_i - 
{\omega_i}_\eta) 
e^{-\int \frac{\eta_i^2}{4\gamma T} d\tau'}  
\end{equation}

where ${\cal N}$ is a normalization constant, and we have introduced an 
external source $J_i(t)$  
which probes the fluctuations of the ergodic ensemble. At the end of 
the calculations, in the thermodynamic limit $N\rightarrow \infty$, 
 $J_i$ would be set to zero. ${\omega_i}_\eta$ is 
the solution of the Langevin equation (1) with the initial probability 
distribution $P(\omega(0))$. \\  
Using some algebraic manipulations we now rewrite this partition 
function in such a manner that the dynamical evolution of the network 
configuration in time becomes more apparent and it is seen that for 
large values of the time, the patterns evolve into the configuration  
corresponding to the ground state energy of the system. \\ 

From (1), we can write :\\ 
\begin{equation} 
\delta(\omega_i - {\omega_i}_\eta) = \delta \left( \dot{\omega_i} + 
\gamma T\frac{\delta E} 
{\delta\omega_i} - \eta_i \right) ~ \left\|\frac{\delta\eta_i}{\delta\omega_i}
\right\| 
\end{equation} 
where the Jacobian $\left\|\frac{\delta\eta}{\delta\omega}\right\|$ of the 
transformation $\eta_i\rightarrow\omega_i$ can be written as : \\  
\begin{eqnarray}
\left\|\frac{\delta\eta_i}{\delta\omega_j}\right\| & = & \det \left[ 
\left( \delta_{ij}\partial_\tau + 
\gamma T\frac{\delta^2 E}  
{\delta\omega_i(\tau)\delta\omega_j(\tau')} \right) 
\delta(\tau - \tau') \right] \nonumber \\ 
& = & \exp \left[{\rm tr} \ln \partial_\tau \left( \delta_{ij}
\delta(\tau - \tau') + {\partial_{\tau'}}
^{-1} \gamma T\frac{\delta^2 E}   
{\delta\omega_i(\tau)\delta\omega_j(\tau')} \right) \right]
\end{eqnarray} 
and ${\partial_{\tau'}}^{-1}$  satisfies :\\ 
\begin{equation} 
\partial_\tau G(\tau - \tau') = \delta(\tau - \tau') 
\end{equation} 
Equation (4) then reduces to :\\
\begin{equation} 
\left\|\frac{\delta\eta_i}{\delta\omega_i}\right\| =  \exp \left\{{\rm tr} 
\left[ 
\ln \partial_\tau + \ln \left( \delta(\tau - \tau') + G_{ij}(\tau - \tau')
\gamma T\frac{\delta^2 E}{\delta\omega_j(\tau)\delta\omega_i(\tau')} 
\right) \right]
\right\} 
\end{equation} 

Since we are primarily interested in the dynamical evolution of the 
patterns of  
the neural network forward in time, the Green's function $G(\tau-\tau')$ must 
satisfy \\
\begin{equation} 
G(\tau-\tau') = \theta(\tau-\tau') 
\end{equation}  

Substituting for $G$ from (7) into (6), and expanding out the logarithm 
we obtain \\ 
\begin{equation} 
\left\|\frac{\delta\eta_i}{\delta\omega_i}\right\| =  e^{\gamma T \int_0^\tau 
d\tau' \theta(0) \frac{\delta^2E}{\delta{\omega_i(\tau')}^2} } 
\end{equation} 
where an overall factor of ${\rm tr}\ln \partial_\tau$ in the exponential 
has been absorbed in the normalization. We choose to work with the  
mid-point prescription $\theta(0) = 1/2$ which leads to \\ 
\begin{equation} 
\left\|\frac{\delta\eta_i}{\delta\omega_i}\right\| =  e^{\frac{\gamma T}{2}
 \int_0^\tau  
d\tau' \frac{\delta^2E}{\delta{\omega_i(\tau')}^2} }  
\end{equation}  
From (1), (3) and (10) the partition function can be rewritten as \\
\begin{eqnarray}
Z[J] & = & {\cal N} \prod_{i,\mu} \int D\omega_i P(\omega(0)) e^{-\frac{1}
{\gamma T} \int_0^\tau 
J_i(\tau')\omega_i(\tau') d\tau'} e^{\frac{\gamma T}{2}\int_0^\tau d\tau' \frac
{\delta^2E}{\delta\omega_i^2}}e^{-\frac{1}{4\gamma T}\int_0^\tau d\tau' 
{(\dot{\omega_i} 
+ \gamma T\frac{\delta E}{\delta\omega_i})}^2 } \nonumber \\  
& = & {\cal N} \prod_{i,\mu} \int D\omega_i P(\omega(0)) e^{-\int_0^\tau
d\tau' \{ \frac{1}{\gamma T} J_i(\tau')\omega_i(\tau') + \frac{1}{4\gamma T}
[ {\dot {\omega_i}}^2 + 
{(\gamma T)}^2{(\frac{\delta E}{\delta\omega_i})}^2 - 2{(\gamma T)}^2 
\frac{\delta^2E}
{\delta\omega_i^2} ] \}} e^{-\frac{1}{2}E(\tau)} e^{\frac{1}{2}E(0)~~~}~~~   
\end{eqnarray} 
The $N$-point correlations between the synaptic efficacies can be calculated 
by taking the $N$-th functional derivative of $Z[J]$ with respect to $J$. 
After making a change of scale of the time variable 
$\tau$ as : ~~$\tau' \rightarrow \tau''= 2\tau'$ we find that the partition 
function can be rewritten as \\
\begin{equation} 
Z[J] = {\cal N} \prod_{i,\mu} \int D\omega_i P(\omega(0)) e^{-\frac{1}{\gamma T}
\int_0^{2\tau}
{\cal L}^{FP}d\tau' - \int_0^{2\tau}d\tau''\frac{1}{\gamma T} 
J_i(\tau'')\omega_i(\tau'')}
e^{-\frac{E(2\tau)}{2}} 
e^{\frac{E(0)}{2}}
\end{equation}
where we have defined a Fokker-Planck lagrangian as :\\ 
\begin{equation} 
{\cal L}^{FP} = \frac{1}{2}{(\frac{d\omega_i}{d\tau})}^2 + 
\frac{{(\gamma T)}^2}{8}
{(\frac{\delta E}{\delta\omega_i})}^2 - \frac{\gamma T}{4}\frac{\delta^2E}
{\delta\omega_i^2}
\end{equation} 
It may be observed that this can be derived from a Fokker-Planck Hamiltonian $H^{FP}$ 
which obeys 
the heat equation \\ 
\begin{equation} 
\frac{\partial\Psi}{\partial\tau} = -2H^{FP}\Psi 
\end{equation}  
where\\
\begin{equation} 
\Psi = e^{\frac{E(\omega)}{2}} P(\omega,\tau) 
\end{equation} 
and\\ 
\begin{equation} 
H^{FP} = \gamma T [ -\frac{1}{2}\frac{\delta^2}{\delta\omega_i^2} + 
\frac{1}{8}{(\frac{\delta E}{\delta\omega_i})}^2 -\frac{1}{4}
\frac{\delta^2E}{\delta\omega_i^2} ]
\end{equation}
It is possible to find a series solution to (14) : \\
\begin{equation} 
\Psi(\omega,\tau) = \sum_n c_n\psi_n e^{-2E_n\tau}  
\end{equation} 
where $c_n$ are normalizing constants and $E_n \ge 0$ are the 
energy eigenvalues 
of the operator equation:\\
\begin{equation} 
H^{FP} \psi_n = E_n \psi_n 
\end {equation}  
It is easy to see that $H^{FP}$ is a positive semi-definite operator 
with its ground state $E_0 = 0$ defined by $\psi_0 = e^{-\frac{E(
\omega)}{2}}$. In the equilibrium limit $t\rightarrow \infty$, only 
the ground state configuration $\psi_0$ contributes to $P(\omega,\tau)$
and we have \\
\begin{equation} 
\lim_{t \rightarrow \infty} P(\omega, \tau) = c_0 e^{-E(\omega)} 
\end{equation}
Interestingly, following Gozzi [9], it is also possible to define a 
canonical momentum ${\hat \Pi}$ conjugate to the variable representing 
the synaptic efficacy,
from the Fokker-Planck lagrangian :\\ 
\begin{equation} 
{\hat \Pi_i} = \frac{\delta{\cal L}^{FP}}{\delta {\dot \omega_i}} = \frac{1}{2\gamma T}
({\dot\omega_i} + \gamma T\frac{\delta E}{\delta\omega_i}) 
\end{equation}
so that the partition function can be written in the form of the Gibbs 
average of equilibrium statistical mechanics :\\
\begin{equation} 
Z[J] = {\cal N}\prod_{i,\mu} \int {\cal D}\omega_i(0){\cal D}
{\hat \Pi_i}(0) e^{- 
\int \{ \frac{1}{2}{\hat \Pi_i}^2 + E(\omega) + 
J_i(\tau')\omega_i(\tau')\}d\tau' } 
\end{equation} 
The advantage of writing $Z$ in this form is that the 
integration measure and $Z$ 
are independent of $\tau$. \\ 
The correlations between the synaptic strengths can be calculated either using  
eqns.(2) within the Langevin approach, or using the equivalent 
Fokker-Planck equations.  
Since the noise-sources are delta-correlated, the correlations between 
the synaptic efficacies would be stationary for a proper choice of the initial 
probability $P(\omega(0))$ :  \\ 
\begin{equation} 
{\langle \omega_\eta(\tau_1) \dots \omega_\eta(\tau_l)\rangle}_
{\eta,P(\omega(0))} 
= \delta (\tau_1-\tau_2, \tau_2-\tau_3, \dots , \tau_l-\tau_{l-1}) 
\end{equation}
where $\langle . \rangle_{\eta, P(\omega(0))}$ denotes average over both $\eta$ 
and $\omega(0)$. 
Since the correlations depend only on the time differences, these would remain 
invariant under a uniform translation in the time :\\ 
\begin{equation} 
{\langle \omega_\eta(\tau_1) \dots \omega_\eta(\tau_l)\rangle}_{\eta,
P(\omega(0))} 
= {\langle \omega_\eta(\tau_1+t) \dots \omega_\eta(\tau_l+t)\rangle}_
{\eta,P(\omega(0))} 
\end{equation}
Setting all $\tau_i$s  equal and taking the $t\rightarrow \infty$ on both sides 
we find that while the left hand side of equation (23) is independent of $t$, 
the right hand side is the average with the equilibrium distribution (19). 
From here it is clear that it is not necessary to take the $t \rightarrow 
\infty$ limit to get the equilibrium steady state distribution --- at every 
finite $t$, the stochastic correlations are already the steady-state 
equilibrium ones. 
As the steady state equilibrium distribution 
for the neural network system corresponds to its lowest energy state, it is 
clear that as $t \rightarrow \infty$ , the patterns evolve into that having 
the lowest energy which acts as the attractor state for the dynamical system.\\

It is necessary to determine the global minimum of the cost function. 
We do this 
by constructing an effective cost function $\Gamma_{eff}[\bar \omega]$ 
by performing a Legendre 
transformation which removes the dependence on $J_i$ in favour of a 
dependence  
on the average $\bar \omega_i$ of the synaptic strengths :\\ 
\begin{equation} 
\Gamma_{eff}[\bar\omega] = -\gamma T \ln Z[J] - 
\int_0^\tau J_i(\tau'){\bar\omega_i}
(\tau') d\tau' 
\end{equation}  
where \\
\begin{eqnarray} 
\bar\omega_i = \langle \omega_i \rangle &=& -\gamma T\frac{\delta 
\ln Z[J]}{\delta J_i} 
\nonumber\\ 
J_i &=& -\frac{\delta\Gamma_{eff}[\bar\omega]}{\delta{\bar\omega_i}}
\end{eqnarray} 
{\em By construction}, the exact effective cost function (24) is 
convex [10] and 
its global minimum gives the ground state energy 
of the neural network system.\\ 
Equations (24) and (25) can be combined to give \\ 
\begin{equation} 
\Gamma_{eff}[\bar\omega] = -\gamma T \ln \prod_{i,\mu} 
\int {\cal D}\omega_i(0) 
e^{\frac{-E(\omega(0))}{2}}{\cal D}\omega_i(\tau) 
e^{-\frac{E(\omega(\tau))}{2}}
{\tilde {\cal D}}\omega_i e^{-\frac{1}{\gamma T}\int_0^\tau [ {\cal L}^{FP} - 
\frac{\delta \Gamma_{eff}}{\delta \bar\omega_i}(\omega_i - 
\bar\omega_i) ] d\tau'} 
\end{equation}
Here, just as in [7], the time interval between $0$ and $\tau$ has been 
sliced into $N-1$ infinitesimal parts:\\ 
\begin{equation} 
{\tilde {\cal D}}\omega_i = \lim_{N \rightarrow \infty} \prod_{i=1}^{N-1} 
{\cal D}\omega_{\tau_i}    ,    
\end{equation}   
$\omega_{\tau_i}$ being the configuration of the synaptic efficacies at 
time $\tau$.\\   
As it is not possible in general to solve this exactly, we find a solution by 
assuming that we can make an expansion of $\Gamma_{eff}[\bar\omega]$ 
in powers 
of a small parameter which we take as $\gamma T$ :\\ 
\begin{equation}
\Gamma_{eff}[\bar\omega] = \sum_{n=0}^\infty {(\gamma T)}^n \Gamma^{(n)}[\bar
\omega] 
\end{equation} 
We discuss later how one can explicitly determine the value of the 
expansion parameter.\\  
Substituting (28) into (26) and making a Taylor expansion of ${\cal L}^{FP}$ 
about $\bar\omega$ we obtain \\ 
\begin{equation} 
\Gamma_{eff}[\bar\omega] = \sum_{i,\mu} {\cal L}^{FP}(\bar\omega)  -  \gamma T
 \sum_{i,\mu} {\cal N}\ln \int {\cal D}\tilde\omega_i(0) 
e^{-E(\omega(0))}{\cal D}
\tilde\omega_i e^{-\int_0^\tau d\tau' \{ \frac{1}{2}\frac{\delta^2 
{\cal L}^{FP}}
{{\delta\tilde\omega_i}^2}{\tilde\omega_i}^2 + \dots - {(\gamma T)}^{1/2}\frac{\delta 
\Gamma^{(1)}}
{\delta \tilde\omega_i}\tilde\omega_i + O{(\gamma T)}^2 \} } 
\end{equation} 
where we have performed the shift :\\ 
\begin{equation} 
{(\gamma T)}^{1/2}\tilde\omega_i = \omega_i - \bar\omega_i 
\end{equation}
By substituting the expansion (28) on the left hand side of (29), we obtain\\
\begin{eqnarray} 
\Gamma^{(0)}[\bar\omega] &=& \sum_{i,\mu}^{ } 
{\cal L}^{FP}(\bar\omega)\nonumber\\ 
\Gamma^{(1)}[\bar\omega] &=& \frac{1}{2}\sum_{i,\mu} 
\ln {\rm det} \frac{\delta^2 
{\cal L}^{FP}}{\delta{\bar\omega_i}^2} 
\end{eqnarray} 
Thus the global minimum for the neural network system can be found by minimising \\
\begin{equation} 
\Gamma_{eff}[\bar\omega] = \sum_{i,\mu} \left[ {\cal L}^{FP}(\bar\omega) 
+ \frac{\gamma T}{2} 
\ln {\rm det}\frac{\delta^2{\cal L}^{FP}}{\delta\bar\omega_i^2} 
+ \cdots \right] 
\end{equation} 
with respect to $\bar\omega_i$. \\ 
In the equilibrium limit the $\dot\omega_i$ term in (13) would not 
contribute to ${\cal L}^{FP}$, so using (13) in (32) and finding the 
root ${\bar{\omega_i}}_{\rm min}$ of the equation \\
\begin{equation} 
\frac{\delta\Gamma_{eff}[\bar\omega]}{\delta\bar\omega_i} = 0 
\end{equation} 
which minimizes $\Gamma_{eff}$, it is possible to avoid the problem of 
spurious minima for any specific choice of the cost function for a 
particular learning rule.\\ 
In this framework, the roots of (33) giving rise to the various local minima  
for the particular neural network under consideration are the source of the 
spurious minima for the system. 
Since $\Gamma_{eff}[\bar\omega]$ is convex, it is clear that by finding the root which gives the minimal value 
of the effective cost function, one can immediately arrive at the global 
minimum of the system without getting any interference from the spurious 
states.\\
 
As an example, consider the cost function for a perceptron 
discussed in [6] : \\ 
\begin{equation} 
E= \frac{1}{2}\sum_\mu {(\zeta^\mu - \frac{1}{\sqrt N}\sum_j 
\omega_j\xi^\mu_j)}^2 + \frac{\lambda}{2}\sum_j{\omega_j}^2 
\end{equation} 
where the constant $\lambda$ was added to keep the connections from going to 
infinity. The external source $J_i$ in our framework plays the role of the 
auxiliary field $h_i$ in [6].\\ 
The Fokker-Planck lagrangian in this case is :\\ 
\begin{equation}
{\cal L}^{FP} = \frac{1}{2}{\dot \omega_i}^2 + \frac{{(\gamma T)}^2}{8} 
{\left( \left(\lambda + \frac{1}{N}\sum_{\mu\nu}\sum_i\xi^\mu_i 
\xi^\nu_i \right)\omega_i - \sum_{\mu\nu}\frac{1}{\sqrt N}\zeta^\mu \xi^\nu_i 
\right)}^2 - \frac{\gamma T}{4} (\frac{1}{N}\sum_{\mu\nu}\xi^\nu_i 
\xi^\mu_i + \lambda) 
\end{equation} 
Substituting ${\cal L}^{FP}$ from (35) into (12) and using the first relation 
in (25) we obtain after setting $J=0$ in the equilibrium limit,\\ 
\begin{equation}
\bar\omega_i = \frac{1}{\sqrt N} \xi^\nu_i {\left(\lambda + 
\frac{1}{N}\sum_{\mu\nu} 
\sum_i \xi^\mu_i \xi^\nu_i \right)}^{-1}_{\nu\mu}\zeta^\mu 
\end{equation} 
This is in agreement with the result obtained in [6].\\ 
From the fluctuation - response theorem, the response function at 
equilibrium is just the full connected propagator $G_{ik}$ which 
is given by :\\ 
\begin{equation} 
G^{-1}_{ik} = {G_0^{-1}}_{ik} + \Sigma_{ik} 
\end{equation} 
where $G_0$ is the tree-level propagator and one can calculate the 
self-energy $\Sigma$ using diagrammatic methods [5,6]. It was shown 
in [5,6] that the 
self energy is given by :\\
\begin{equation} 
\Sigma = \frac{\alpha}{1+G} 
\end{equation} 
where  $\alpha = p_{\rm max}/N$ is the storage capacity of the network. 
In these papers the authors calculated the storage capacity of the 
network at equilibrium using diagrammatic methods. \\  
Since we have  $G^{-1}_{ij}= {(\gamma T)}^{-1}\frac{\delta^2\Gamma_{eff}}
{\delta\bar\omega_i \delta\bar\omega_j}$ , 
we can also 
determine the storage capacity analytically from : \\ 
\begin{equation} 
\alpha = 1 + {(\gamma T)}^{-1}\frac{\delta^2\Gamma_{eff}}{\delta
{\bar \omega}^2} - 
\frac{\delta^2{\cal L}^{FP}}
{\delta {\bar \omega}^2}\left( 1 + \gamma T{\left(\frac{\delta^2\Gamma_{eff}}
{\delta {\bar \omega}^2}\right)}
^{-1} \right) 
\end{equation}  
Since we have assumed that ~$\gamma T$ ~ is a small parameter, it is 
sufficient to work upto first order in the $\gamma T$ expansion of 
$\Gamma_{eff}$.  
The steady-state equilibrium limit $\tau \rightarrow \infty$  
corresponds to the most 
ordered state when the patterns have settled into their attractor states. 
In the phase space of synaptic strengths, it is thus a 
state of minimum symmetry or zero entropy. 
This means that in the equilibrium limit,\\ 
\begin{equation} 
\frac{\delta}{\delta T}\left( -\gamma T \ln Z[J] \right) = 
\frac{\delta}{\delta T}\left( \Gamma_{eff}[\bar\omega] 
+ \frac{1}{\gamma T}\int_0^\tau 
J_i(\tau')\bar\omega_i(\tau') d\tau' \right)  = 0 
\end{equation}
In the thermodynamic limit $N\rightarrow \infty$, $J$ is set to zero, and we 
get the result that for the attractor states:\\
\begin{equation} 
\frac{\delta \Gamma_{eff}[\bar\omega]}{\delta T} = 0 
\end{equation} 
which shows that in this limit the effective cost function is stable with 
respect to changes in the noise level.\\  
Condition (41) can be used to determine the value of the quantity $\gamma T$ 
at which the system settles into an attractor state. 
For the cost function (34) for example, one obtains :\\ 
\begin{equation} 
\gamma T = \frac{1}{\sqrt 3}\frac{{(\frac{1}{N}\sum_{\mu\nu}\xi^\nu_i 
\xi^\mu_i + \lambda )}^{1/2}}{{\left[\ln {\rm det}(\lambda + \frac{1}{N}
\sum_{\mu\nu}\sum_i\xi^\mu_i\xi^\nu_i) \right]}^{1/2}} 
\end{equation}
Applying equations (32) and (39) to the example of (34),and using 
eqn.(35), we obtain  
the following result for $\alpha$ : \\ 
\begin{equation} 
\alpha = (1-\gamma T) \left( 1 + \frac{\gamma T}{4}\sum_i {(\lambda + 
\frac{1}{N}\sum_{\mu\nu}\xi^\mu_i\xi^\nu_i)}^2 \right) 
\end{equation} 
The value of $\gamma T$ obtained in (42) can then be substituted into (43) 
so that in the thermodynamic limit one obtains the result \\ 
\begin{eqnarray} 
\alpha & = & 1 + \frac{1}{\sqrt 3}{\left( \frac{\frac{1}{N}\sum_{\mu\nu}
\xi^\nu_i\xi^\mu_i + \lambda}{\ln {\rm det}(\lambda +\frac{1}{N}
\sum_{\mu\nu}\sum_i\xi^\mu_i\xi^\nu_i)} \right)}^{1/2} 
\left( \frac{1}{4}{(\lambda + \frac{1}{N}\sum_{\mu\nu}\sum_i\xi^\mu_i
\xi^\nu_i)}^2 - 1 \right) \nonumber \\ 
&   &  -\frac{1}{12}\left( \frac{\frac{1}{N}\sum_{\mu\nu}
\xi^\nu_i\xi^\mu_i + \lambda}{\ln {\rm det}(\lambda +\frac{1}{N}
\sum_{\mu\nu}\sum_i\xi^\mu_i\xi^\nu_i)} \right) {(\lambda + \frac{1}{N} 
\sum_{\mu\nu}\sum_i \xi^\mu_i\xi^\nu_i)}^2 
\end{eqnarray} 
for the storage capacity of the network described by the cost 
function (34).\\ 
Although in the pseudo-inverse learning rule considered in the example 
above, the relation between the input and the output is linear, our 
construction of the effective cost function and our method for 
calculating the storage capacity of the network is applicable also for 
models with non-linear input-output relation such as :\\ 
\begin{equation} 
\zeta^\mu = f(\sum_j \omega_j\xi^\mu_j/\sqrt N) 
\end{equation} 
where $f(x)$ is a non-linear function of x.
The procedure itself at no stage depends on any particular model and 
is independent of whether the input-output relation is linear or 
non-linear. \\ 
In this manner it is possible to calculate the storage capacity 
for any neural network system, in a completely analytical manner, 
without having to resort to diagrammatic methods, and independent of 
the choice of the cost function. It may be observed that the procedure 
we have set up in the foregoing allows the explicit calculation of the 
synaptic efficacies $\omega_{ij}$ from the input $\xi^\mu_i$ and the 
output $\zeta^\mu_i$.  \\

\subsection*{Discussion} 

We have shown that a neural network system can be viewed as a non-equilibrium 
stochastic system of synaptic efficacies which evolve for very large values 
of the time into an equilibrium configuration having the lowest energy 
which acts as the attractor state of the network. \\
An effective cost function is constructed and a perturbative scheme is 
developed for calculating it. The global minimum of the effective cost 
function can be determined and gives the exact ground state energy 
of the system. It is shown that in the thermodynamic limit, the effective 
cost function is invariant under changes in the noise level. In the 
perturbation expansion we have constructed for $\Gamma_{eff}$, we have 
assumed that the expansion parameter $\gamma T$ is small. This can be 
ensured by always keeping the noise-level low so that $\gamma T \ll 1$.\\  
In this paper we have constructed a path integral framework for the 
simplest case of a perceptron. It should be possible to generalize 
this construction for the more realistic case of many-layered 
neural networks and to arrive at their global minima straightaway 
without having to bother about the various local minima. \\

\subsection*{Acknowledgement} 
I would like to acknowledge support from the Institute for Robotics \& Intelligent 
Systems, Centre for Artificial Intelligence \& Robotics, Bangalore, India, 
during a large part of this work. 
\newpage 
\subsection*{References} 
\begin{enumerate}
\item W.A.Little, Math.Biosci.{\bf 19}, 101 (1974).
\item E.Gardner, J.Phys.A {\bf 21}, 257 (1988).
\item E.Gardner \& B.Derrida, J.Phys.A {\bf 21}, 271 (1988). 
\item B.Muller \& J.Reinhardt, {\em Neural Networks : An Introduction}, 
Springer-Verlag,(1990). 
\item J.A.Hertz \& G.I.Thorbergsson, Physica Scripta {\bf T25}, 149 (1989). 
\item J.A.Hertz, A.Krogh \& G.I.Thorbergsson, J.Phys.A {\bf 22}, 2133 (1989). 
\item H.Sompolinsky \& A.Zippelius, Phys.Rev.{\bf B25}, 6860 (1982).\\
A.Crisanti \& H.Sompolinsky, Phys.Rev.{\bf A36}, 4922 (1987).\\ 
R.Kree \& A.Zippelius, Phys.Rev.{\bf A36}, 4421 (1987).\\                             
H.Sompolinsky, N.Tishby \& H.S.Seung, Phys.Rev.Lett.{\bf 65}, 1683 (1990). 
\item E.Gozzi, Phys.Rev.{\bf D28}, 1922 (1983).
\item E.Gozzi, Phys.Lett.{\bf 130B}, 183 (1983). 
\item J.Iliopoulos, C.Itzykson \& A.Martin, Rev.Mod.Phys.{\bf 47}, 165 
(1975).                              
\end{enumerate}

\end{document}